\documentclass[11pt,preprint]{aastex}
\usepackage{times}
\usepackage{graphicx}
\let\oldfootsep=\footnotesep
\newcommand\ltsima{$\; \buildrel <\over\sim \;$}
\newcommand\simlt{\lower.5ex\hbox{\ltsima}}
\newcommand\gtsima{$\; \buildrel >\over\sim \;$}
\newcommand\simgt{\lower.5ex\hbox{\gtsima}}
\newcommand\etal{et~al.}

\newcommand\msun { \rm {M_\odot}}

\newcommand\pac{Paczy{\'n}ski }

\shorttitle{}
\shortauthors{Bennett, Becker and Tomaney}

\begin{document}


\title{Photometric Confirmation of MACHO 
         Large Magellanic Cloud Microlensing Events}


\author{David~P.~Bennett\altaffilmark{1},
               Andrew~C.~Becker\altaffilmark{2}, and
               Austin~Tomaney\altaffilmark{3}
       } 
\altaffiltext{1}{Department of Physics,
    University of Notre Dame, IN 46556, USA\\
    Email: {\tt bennett@nd.edu}}

\altaffiltext{2}{Astronomy Department,
    University of Washington, Seattle, WA 98195, USA\\
    Email: {\tt becker@astro.washington.edu}}

\altaffiltext{3}{Applied Biosystems,
    850 Lincoln Centre Dr., Foster City, CA 94404, USA\\
    Email: {\tt TomaneAB@appliedbiosystems.com}}


\clearpage

\begin{abstract}
We present previously unpublished photometry of three Large 
Magellanic Cloud (LMC) microlensing events and show that the
new photometry confirms the microlensing interpretation of these
events. These events were discovered by the MACHO Project 
alert system and were also recovered by the analysis
of the 5.7 year MACHO data set. This new photometry provides a substantial
increase in the signal-to-noise ratio over the previously published
photometry and in all three cases, the gravitational microlensing interpretation 
of these events is strengthened. The new data consist of MACHO-Global
Microlensing Alert Network (GMAN) follow-up images from 
the CTIO 0.9 telescope plus difference imaging photometry of the 
original MACHO data from the 1.3m ``Great Melbourne" telescope at 
Mt.~Stromlo. 
We also combine microlensing light curve fitting with photometry from 
high resolution HST images of the source stars to provide further
confirmation of these events and to show that the microlensing
interpretation of event MACHO-LMC-23 is questionable. Finally, we
compare our results with the analysis of
Belokurov, Evans \& Le Du who have attempted
to classify candidate microlensing events with a neural network method,
and we find that their results are contradicted by the new data and
more powerful light curve fitting analysis for each of the four events
considered in this paper. The failure of the
Belokurov, Evans \& Le Du method is likely to be due to their use of a set
of insensitive statistics to feed their neural networks.
\end{abstract}




\section{Introduction}
\label{intro}
The decade of the 1990s saw the development of gravitational 
microlensing as a new method to detect objects in the planet
through stellar
mass range through their gravitational effect on light rays from
background stars \citep{macho-nat93,eros93,ogle93}.
Applications of this method include such diverse topics as the
search for baryonic dark matter in the Milky Way halo
\citep{macho-lmc2,macho-lmc5.7,eros-lmc-tau,eros-smc-tau}, the
discovery of planets orbiting distant stars \citep{bond-moa53},
limb darkening measurements of distant stars
\citep{planet-97blg28}, and obtaining high S/N spectra of distant
stars that would otherwise be possible only with future
extremely large telescopes \citep{lennon-96blg3,minniti-blg-li}.

The original proposal for this method was the suggestion by
\citet{pac86} to use microlensing to determine if the Milky Way's
dark halo is made up of brown dwarfs or jupiters. (Dark matter
objects that can be detected via microlensing are often referred to as
MAssive Compact Halo Object or MACHOs.) The MACHO and
EROS groups have completed the microlensing survey proposed
by \citet{pac86}, and their main result is the conclusion that
the Milky Way's dark halo is {\it not} dominated by objects
with masses in the planet-stellar mass range 
\citep{macho-eros-spike,macho-lmc2,macho-lmc5.7,eros-lmc-tau,macho-30msun,eros-smc-tau}.
Objects in the entire mass range, $10^{-7} - 30\,\msun$ are excluded 
from dominating the Milky Way's dark halo, with upper limits extending
down to $< 5$\% of the dark halo mass for sub-stellar mass objects.
However, the MACHO Project \citep{macho-lmc5.7} has found a microlensing
signal above the expected background due to lensing by known
stellar populations in the Milky Way and Large Magellanic
Cloud (LMC).  This signal has
a microlensing optical depth equivalent to a dark halo with a
MACHO fraction of about 20\%, although the total mass could
be somewhat smaller with if the MACHOs have a distribution
that resembles a spheroid or a very thick disk rather
than the dark halo \citep{gates-gyuk-disk}. Some of the first results
from M31 microlensing observations appears to confirm this result
\citep{m31-halolens,m31-tau}.
The timescale of the LMC microlensing events
indicates a typical mass of $\sim 0.5 \msun$, so the most 
plausible lens objects are white dwarfs, because main sequence
stars of this mass would be much too bright to have previously
escaped detection. However, there are a number of potential
problems with the white dwarf explanation of this LMC microlensing excess 
\citep{torres-opp-WD,flynn-opp-WD,brook-WDhalosim,MC_halo_WD,diskorhalo_WD}.
On the other hand, many of the constraints are evaded if most of the halo white 
dwarfs have Helium atmospheres. It might also be possible to explain
the MACHO results with a population dominated by lower mass
objects \citep{rahvar05}.

The leading alternative explanation for the LMC microlensing excess is that
the lens objects are ordinary stars associated with the LMC 
\citep{sahu94}. However, standard models of the LMC predict that MACHO should
have detected only 2-4 events 
from known stellar populations \citep{wu94,macho-lmc5.7},
and there is a simple dynamical explanation for the small LMC self-lensing
optical depth \citep{gould-disktau}. 
On the other hand, the LMC is not an isolated
galaxy and may have had significant dynamical disturbances from the 
Milky Way \citep{weinberg,evans-kerins}, but current LMC models that include these effects still
cannot account for the observed microlensing events 
\citep{gyuk-lmc-mod,alves-lmc-mod,man-lmc-mod,nikolaev-lmc-mod} 
because most of the events do not occur in the regions of the highest 
predicted LMC self-lensing rate.

The other logical possibility is that the background of microlensing events due
to faint stars in the local Galactic disk could have been underestimated, but 
the lens stars for such events should be readily observed in both high resolution 
and IR observations as event MACHO-LMC-5 has demonstrated 
\citep{macho-hstlmc5,dck-lmc5,lmc5-mass,spitz-lmc5} In fact,
most candidate Milky Way disk events are readily identifiable through their 
anomalous unmagnified colors in the MACHO data, and of the other published
microlensing candidates, only MACHO-LMC-20 appears consistent with
lensing by a low mass disk star. Thus, this possibility 
appears to be less likely than halo or LMC lenses.

Of course, all of this assumes that the MACHO LMC microlensing results
are correct. The EROS collaboration has also monitored the LMC for
microlensing events, but have detected fewer events than MACHO and
reported an upper limit on the microlensing optical depth that is only
barely consistent with the MACHO detection
\citep{eros-lmc-tau}. A similar near-discrepancy
holds for microlensing towards the Galactic bulge, where MACHO also
measured a higher microlensing optical depth \citep{pop-blgclump}
than EROS \citep{eros-blg-tau} In this case, the MACHO result is clearly 
favored as it is more consistent with separate measurements by 
OGLE \citep{ogle-tau} and MOA \citep{moa-tau}, as well as a completely 
independent measurement by MACHO \citep{macho-bulge-diff}.

\subsection{Event Selection by Neural Networks}
\label{nnet}

The difficulty with interpreting the LMC results combined with the 
near-discrepancy between the MACHO and EROS results, has led 
\citet{BEL-blg,BEL-lmc} (hereafter BEL) to attempt to
develop an independent microlensing event selection method for the MACHO
data set. In an attempt to improve upon the MACHO event selection method,
BEL have introduced a method using neural networks instead of the series
of simple cuts on statistics used by MACHO. 
This method seemed promising because neural
networks should be able to find a set of event selection cuts that
are more efficient at identifying microlensing events than can be easily
obtained by more traditional trial-and-error methods. However, a major
drawback of neural networks is their ``black box" nature: it is not obvious
exactly why the neural network might make its event classification decisions.
Thus, a neural network could mysteriously fail to identify real microlensing
events if there was some subtle difference between the real data and the
example events used in the training set for the neural network, and it 
would be difficult to find such an error. This is the reason that 
neural networks were not used in the MACHO analysis, and it appears
to have led BEL and \citet{evans-crap} to over--interpret their results.

In fact, the specific neural network implementation of BEL has a number
of serious flaws. Ideally, one would simply feed the raw time series 
photometry and error estimates into the neural network, and let the neural 
network sort out how to find microlensing events. With hundreds or 
thousands of input values, such a neural network would be far too slow to 
be practical. A practical neural network implementation can be obtained 
if we reduce the raw light curve data to a much smaller number of 
statistics that can be calculated from the raw data. The choice of
these statistics is, of course, critical to the success of the neural 
network classification method, and this is where the method of BEL 
appears to fail. BEL select a set of statistics that are based upon 
auto- and cross-correlation functions of the light curve data, with 
the time series treated as if they were sampled uniformly in time 
(which they certainly are not). The BEL statistics appear to ignore 
the measurement errors and to ignore much of the time history information, 
so it would be somewhat surprising if it could do as well as the MACHO 
selection method. 

Finally, \citet{griest-thomas} have identified a serious logical flaw in the
argument of \citet{evans-crap} who claim that the LMC microlensing
optical depth must be lower than the MACHO value. 
The BEL analysis cannot seriously be considered as a 
method to select microlensing events from a complete data set of millions
of light curves, because its false alarm rate is 1 in $10^4$, which is much
higher than the microlensing rate for any plausible Galactic model. Thus,
BEL use their method in the only way that they can, they apply it to events
that have already passed the MACHO selection criteria. Thus, there is no chance
that they could detect events that MACHO missed, so their detection
efficiency is necessarily lower, contrary to the claim of \citet{evans-crap}.
As we shall see in this paper, the BEL method fails to confirm 3 events that
are almost certainly microlensing events, while ``confirming" one 
microlensing candidate that is almost certainly a variable star. Thus,
it seems clear that the current version of the BEL analysis does not
help to identify actual microlensing events.

\subsection{Paper Organization}
\label{paper_org}

This paper is organized as follows. Sec.~\ref{data_phot} describes the
data sets used for our analysis and the photometry codes used to
convert the images into photometric measurements. Sec.~\ref{lc_fit}
describes our method for light curve fitting and comparison with HST
photometry of the candidate microlensed source stars. The detailed
modeling of the 4 individual events presented in this paper are
presented in sub-sections \ref{event-lmc4}--\ref{event-lmc23}. A
detailed comparison of the results of BEL's analysis with our
results and other additional data is presented in Sec.~\ref{confront},
and in Sec.~\ref{conclude}, we discuss the implications of our
results for the interpretation of the microlensing results towards 
the LMC.

\section{Data and Photometry}
\label{data_phot}

\subsection{MACHO Data}
\label{macho_data}

The events presented in this paper were all identified as microlensing
candidates in \citet{macho-lmc5.7}, based upon photometry with the
standard MACHO SoDophot photometry routine. The SoDophot
photometry presented in this paper is essentially identical to that
presented in \citet{macho-lmc5.7}. A subset of the MACHO images
for these events were
also reduced with the DifImPhot difference imaging photometry
package \citep{tomaney-dip}. In very crowded star fields, the difference
imaging method is usually more accurate than the profile fitting method 
employed by SoDophot \citep{macho-dia1, macho-bulge-diff, ogle-dia-ev}. 
Also, the SoDophot and DifImPhot photometry codes are likely to produce 
different systematic errors, so a comparison of SoDophot and DifImPhot
photometry of the same images can reveal data points that might have
be affected by these systematic errors.

\subsection{Microlensing Alerts and CTIO
 Follow-up Data}
\label{ctio_data}

Event MACHO-LMC-4 was the first LMC microlensing event
discovered and announced in progress, and this occurred on 1994, Oct. 14.
It was discovered while staffing arrangements for MACHO/GMAN service observing
on the CTIO
\footnote [4]
{Cerro Tololo Inter-American Observatory, National Optical
Astronomy Observatories, operated by the Association of Universities for
Research in Astronomy, Inc., under cooperative agreement with the
National Science Foundation.}
0.9 m telescope were still being finalized, and as a consequence,
it was not possible to take CTIO data through the peak of the event. Because of
this problem with the CTIO data, extra observations were scheduled with the
MACHO survey telescope (the 1.3m ``Great Melbourne" telescope at Mt. Stromlo)
until observations from CTIO could resume. The CTIO data set for this 
consists of a single observation before the light curve peak, taken shortly 
after the event was discovered, and 155 observations taken after the light 
curve peak, including 10 baseline observations taken more than a year after 
the peak. These CTIO data were reduced with the ALLFRAME package 
\citep{allframe}, and 0.5\% was added in quadrature to the ALLFRAME error 
estimates to account for
low level systematic errors such as flatfielding errors. (For the MACHO
SoDophot photometry, 1.4\% is added in quadrature to the SoDophot output
errors as described in \citet{macho-lmc5.7}.) 134 of these 156 measurements
pass our data quality cuts \citep{macho-alert} and are included in this analysis. This 
is in addition to 730 MACHO-red band and 728 MACHO-blue band photometric
measurements from SoDophot which pass our data quality cuts, as well as
85 MACHO-red band and 78 MACHO-blue band images reduced with DifImPhot.
The images reduced with DifImPhot were chosen to cover the region of the
microlensing magnification as well as a bit of the baseline.

Microlensing event MACHO-LMC-13 was detected and announced on 1996, Feb. 11,
near peak magnification as MACHO 96-LMC-1. Additional observations 
were immediately
requested from the CTIO 0.9m, and 376 CTIO R-band images were eventually
obtained including 145 baseline measurements taken more than 180 days 
after the light curve peak. 332 of these CTIO measurements pass our
data quality cuts. The data from the MACHO survey includes 1074
MACHO-red measurements and 1176 MACHO-blue measurements reduced
with SoDophot as well as 84 measurements covering the magnified part
of the light curve in each MACHO band reduced with DifImPhot. In addition,
there were 47 R-band and 64 V-band measurements from the UTSO 0.6m telescope. 
These are included in our fits, but they have only limited time coverage
and have larger error bars than the CTIO measurements, and so they are not 
included in our figure. These data were reduced in the same way as the
CTIO and LMC-4 data.

The final event in our sample is MACHO-LMC-15 which was detected and
announced on 1997, Jan. 15 as MACHO 97-LMC-1, well
before peak magnification. Additional observations were again immediately
requested from the CTIO 0.9m, and 74 CTIO R-band images were obtained,
including 26 baseline measurements taken more than 180 days after the
light curve peak. 60 of these CTIO observations pass our data quality cuts.
The MACHO survey data set consists of 475 MACHO-red and 586
MACHO-blue SoDophot measurements as well as 47 MACHO-red
and 49 MACHO-blue DifImPhot measurements. The DifImphot data
covers the magnified portion of the light curve along with the baseline
immediately before and after the event.

Important constraints on our analysis of the light curve
data come from the high resolution HST WFPC2 images of the
microlensed source stars after the microlensing events were
over. The reduction of the HST data and the identification of the
microlensed source stars are discussed in \citep{macho-hstlmc1,macho-hstlmc2}.

\section{Testing the Microlensing Hypothesis with HST Constrained 
Light Curve Fitting}
\label{lc_fit}

Unlike many types of stellar variability, it is possible to calculate 
theoretical microlensing light curves with an accuracy that is much 
better than observational errors, and this makes light curve fitting 
an extremely powerful method for testing the hypothesis that a 
candidate microlensing event was indeed caused by microlensing. The 
light curve fitting method is extremely effective in discriminating against the largest 
background to microlensing events in the LMC, supernovae of type Ia, 
as demonstrated in \citet{macho-lmc5.7}. This success is due to the 
fact that there exist accurate models for type Ia supernova light 
curves. The situation is different for supernovae of type II, which 
do not have an accurate light curve model. The MACHO Project 
\citep{macho-lmc5.7} was also able to exclude its type II SNe background 
by doing fits to SN type 1a light curve models, but this method was 
only convincing because these same events also had detectable galaxy 
hosts visible in the MACHO images or in other ground based images with 
slightly higher resolution. (Since type II SNe occur with young host 
stars and are usually fainter than type Ia SNe, they generally have 
much brighter host galaxies.)

After the removal of the supernovae, the confirmation of microlens 
candidates becomes a bit more difficult because little is known about 
the extremely rare types of variable stars that might mimic microlensing 
events. Thus, we are limited to testing the microlensing model rather 
than comparing it against another model. This task is complicated by the 
fact that the photometric errors are mildly non-Gaussian, although the 
MACHO and GMAN data have managed to remove most of the photometry 
outliers with cuts on a number of data quality flags. One model parameter 
that is particularly sensitive to systematic photometry errors is the 
baseline brightness of the source star. In the crowded fields observed 
by microlensing surveys, it is quite common for the source star to be 
blended with other stars within the same seeing disk. Thus, the source 
star brightness is usually used as a fit parameter, but there is a 
near degeneracy in light curve shapes that can make this parameter 
difficult to determine from a fit.

Another complication is the fact that $\sim 10$\% of microlensing events 
are ``exotic" events that don't follow the standard \pac\ light curve. 
In some cases, such as caustic crossing binary lens events, the ``exotic" 
features are so unique to lensing that there can be no ambiguity in 
the interpretation of the event, such as a caustic crossing binary
lens event like MACHO-LMC-9 \citep{macho-lmc9}.
However, in other cases, such as MACHO-LMC-22 
\citep{macho-lmc5.7}, non--microlensing variability can be fit with 
an exotic lens model.

In view of these considerations, we adopt the following series of steps 
in order to test the microlensing hypothesis for candidate microlensing 
events towards the LMC:
\begin{enumerate}
\item We fit each event with a light curve constraining the MACHO-blue
      source star brightness to match the V magnitude from the HST
      observations. (The tranformation from MACHO-blue to the standard
      Cousins V-band is only weakly dependent on the MACHO-red 
      magnitude \citep{macho-calib}.)
      The best unconstrained fit should not have a fit
      $\chi^2$ that is significantly lower than the constrained 
      fit--unless the source star itself is blended with the lens star.
\item The $\chi^2/{\rm d.o.f.}$ values for the fit should be consistent
      with expectations for real microlensing event. Of particular
      interest are the $\chi^2/{\rm d.o.f.}$ values for the follow-up
      CTIO and DifImPhot data points, and the $\chi^2/{\rm d.o.f.}$
      values in the light curve peak region (defined as the region 
      where the best fit magnification is $A > 1.1$). Comparison of the
      different photometry data sets can indicate if an apparent
      deviation may be due to systematic photometry errors.
\item If a standard \pac\ model is not a good fit to the data, is there
      an exotic model that gives a good fit with plausible parameters?
      If so, are the number of candidates that must be fit with exotic
      models consistent with expectations?
\item The fit MACHO-red baseline magnitude for the source star is
      converted to the standard Cousins R-band and compared to the
      HST R-band magnitude of the source star. If it doesn't match, is
      it consistent with a blending model?
\end{enumerate} 

\begin{figure}
% \plotone{chi2hist.eps}
\plotone{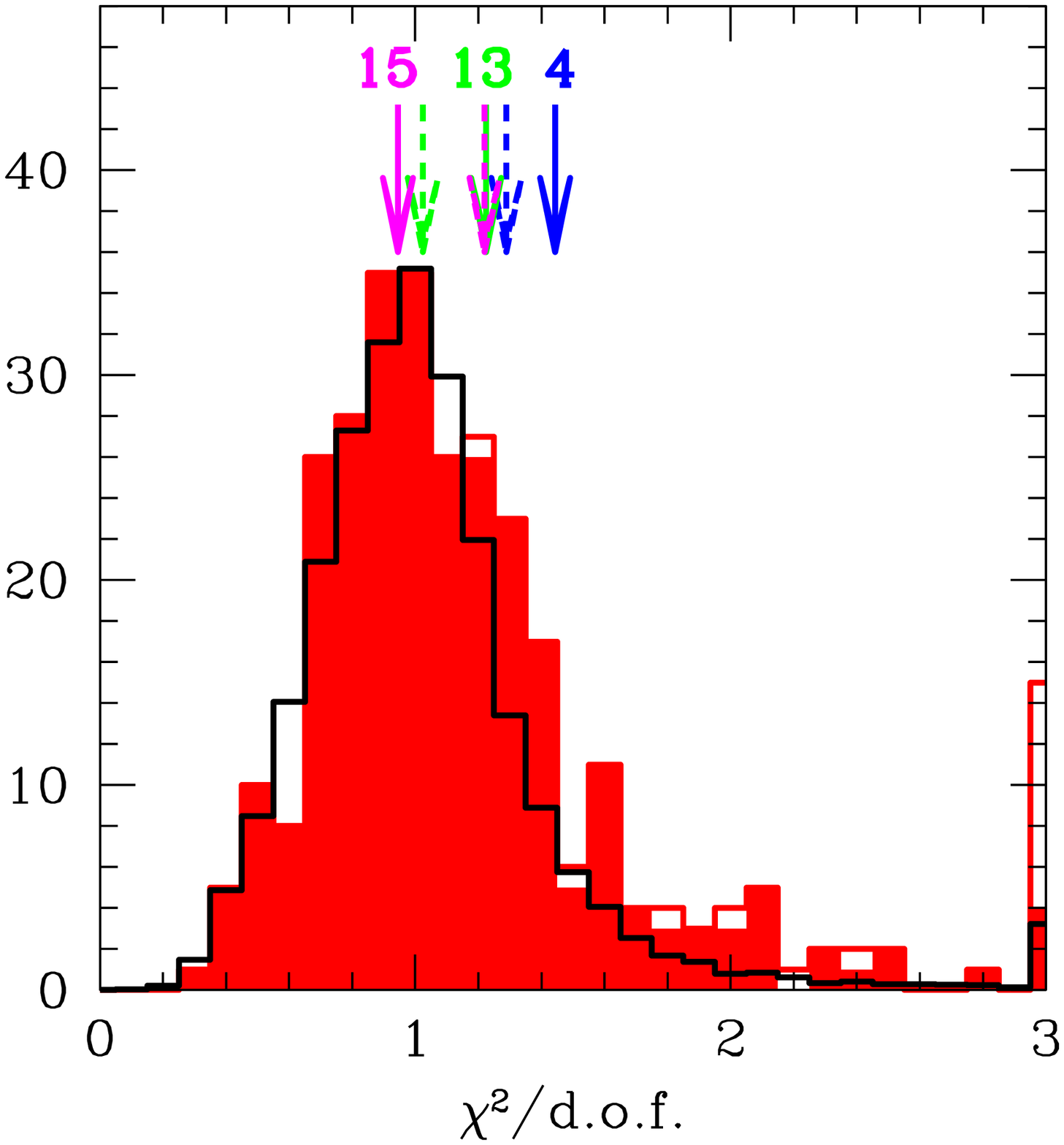}
\caption{
The distribution of reduced $\chi^2/{\rm d.o.f}$ for microlensing light curve 
fits is shown for simulated events in the MACHO LMC data (black 
histogram) and for observed Galactic bulge events (solid red histogram).
The unfilled part of the 
red histogram indicates non-standard microlensing events, such as 
microlensing parallax and binary lens events.
The solid arrows indicate the fit $\chi^2/{\rm d.o.f}$ values for the MACHO
observations of three events presented here: MACHO-LMC-4, 13 and 15.
The dashed arrows indicate the $\chi^2/{\rm d.o.f}$ values for the
additional photometry presented in this paper.
\label{fig-chi2}}
\end{figure}

When making a goodness of fit judgement, it is important to keep in mind
that the photometry errors are non-Gaussian, so Gaussian probabilities
do not apply. Fig.~\ref{fig-chi2} shows the distribution of fit 
$\chi^2/{\rm d.o.f.}$ values for a set of Monte Carlo events from the
LMC efficiency analysis \citep{macho-lmc5.7eff} (black outline histogram)
as well as the observed distribution for Galactic bulge microlensing 
events found by the MACHO alert system \citep{macho-alert} (red histogram).
Events that have been classified as ``exotic" microlensing events can
be expected to have a poor fit to a standard \pac\ light curve, and these
are indicated by the red outline histogram. Note that all the events with
$\chi^2/{\rm d.o.f.} \geq 3$ have been displayed in the 
$\chi^2/{\rm d.o.f.} = 3$ bin. The light curves for the events displayed
in these histogram typically have 500--2000 total data points (including
both MACHO-red and MACHO-blue observations), so the width of the
$\chi^2/{\rm d.o.f.}$ distribution is somewhat larger than we'd expect
from Gaussian statistics. While there is considerable overlap between the
Monte Carlo and actual Galactic bulge microlensing event fit 
$\chi^2/{\rm d.o.f.}$ distributions, the real data distribution extends to
larger values because some types of systematic photometry errors are
not captured by the Monte Carlo simulations.

The fit $\chi^2/{\rm d.o.f.}$ values for the events presented in this paper 
are listed in Table~\ref{tbl-fitpar} and the fit $\chi^2/{\rm d.o.f.}$ values for 
the events we present in this paper, LMC-4, 13, and 15, are in the
82nd, 64th, and 42nd percentiles, respectively, 
of the observed $\chi^2/{\rm d.o.f.}$ distribution
for non-exotic galactic bulge events (the solid red histogram in 
Fig.~\ref{fig-chi2}). In contrast, the dubious microlens candidate, 
LMC-23, is in the 91st percentile (see Sec.~\ref{sec-lmc23}. 
This alone is not enough to 
disqualify this event as microlensing, particularly because a
microlensing parallax model gives a significantly better
$\chi^2/{\rm d.o.f.}$.

Before proceeding to discuss the individual events, we should note
that all the fits presented here are joint fits to the CTIO data as well
as both the SoDophot and DifImPhot reductions of the MACHO
data. This is necessary because only the SoDophot reductions
have been calibrated to the Cousins system and because the
DifImphot photometry of the unmagnified portions of the light
curves are limited. This means that the MACHO images that have
yielded photometry by both methods are used twice, and we must
be careful that this double counting does not affect our results.
In order to correct for the effects of this double counting, we have
adjusted the error bars in Table~\ref{tbl-fitpar} to coincide with the
fractional errors obtained by fitting only the SoDophot+CTIO data
or the DifImPhot+CTIO data (whichever has smaller
error estimates). Also, the total $\Delta\chi^2$ values presented
in Table~\ref{tbl-delchi2} are the sum of the individual $\Delta\chi^2$
values for the CTIO data plus the larger of the SoDophot or the
DifImphot $\Delta\chi^2$ values.

\subsection{Event MACHO LMC-4}
\label{event-lmc4}

\begin{figure}
% \plotone{lclmc4_3plongav8.eps}
\plotone{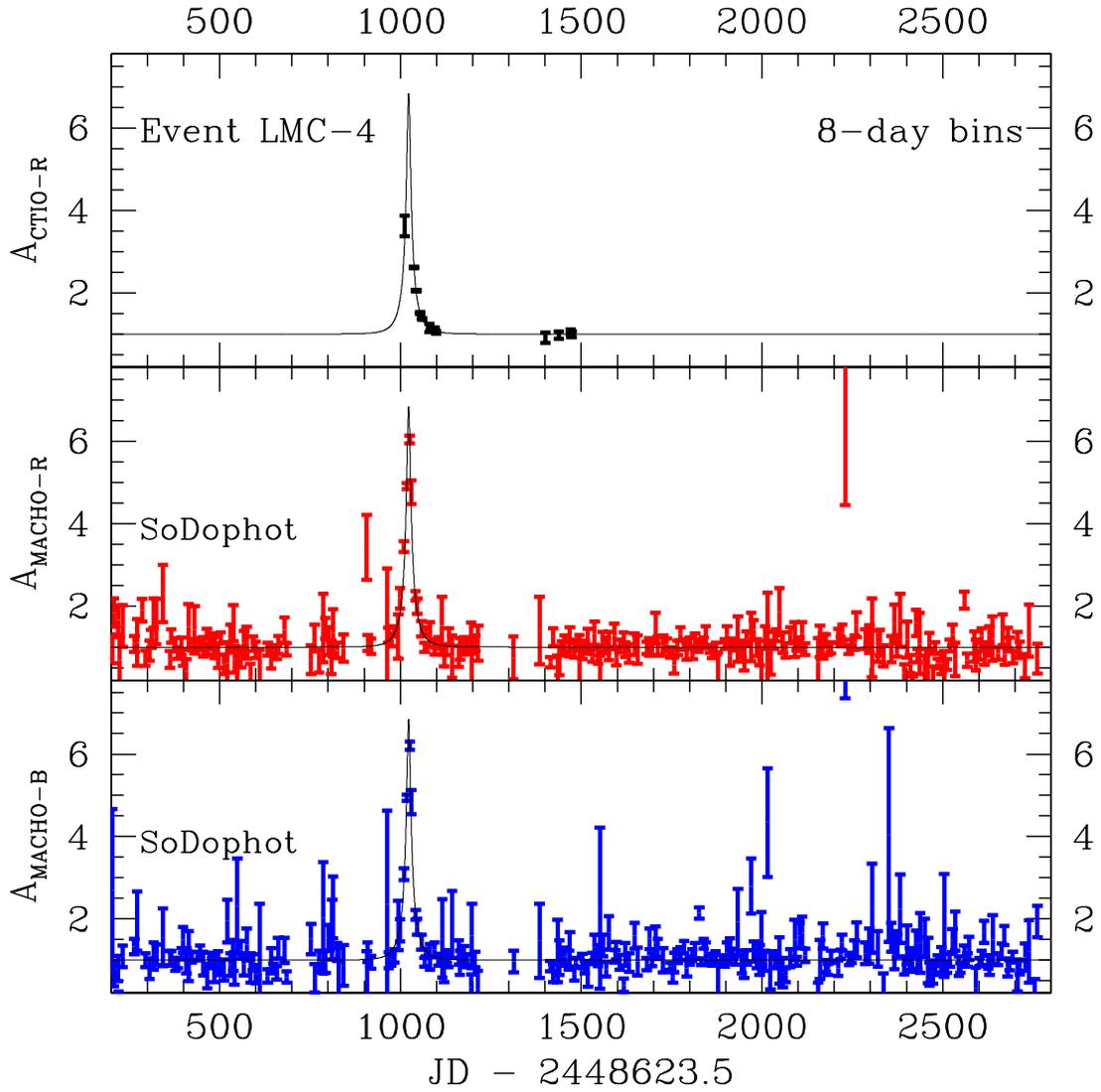}
\caption{
The MACHO and GMAN-CTIO (top panel) follow-up data are shown 
in the full light curve for microlensing event LMC-4. 
The solid line is the best fit light curve constrained to match 
the HST V-band magnitude, and all data is presented in 8-day bins.
\label{fig-lmc4long}}
\end{figure}

\begin{figure}
% \plotone{lclmc4_5pfigav2.eps}
\plotone{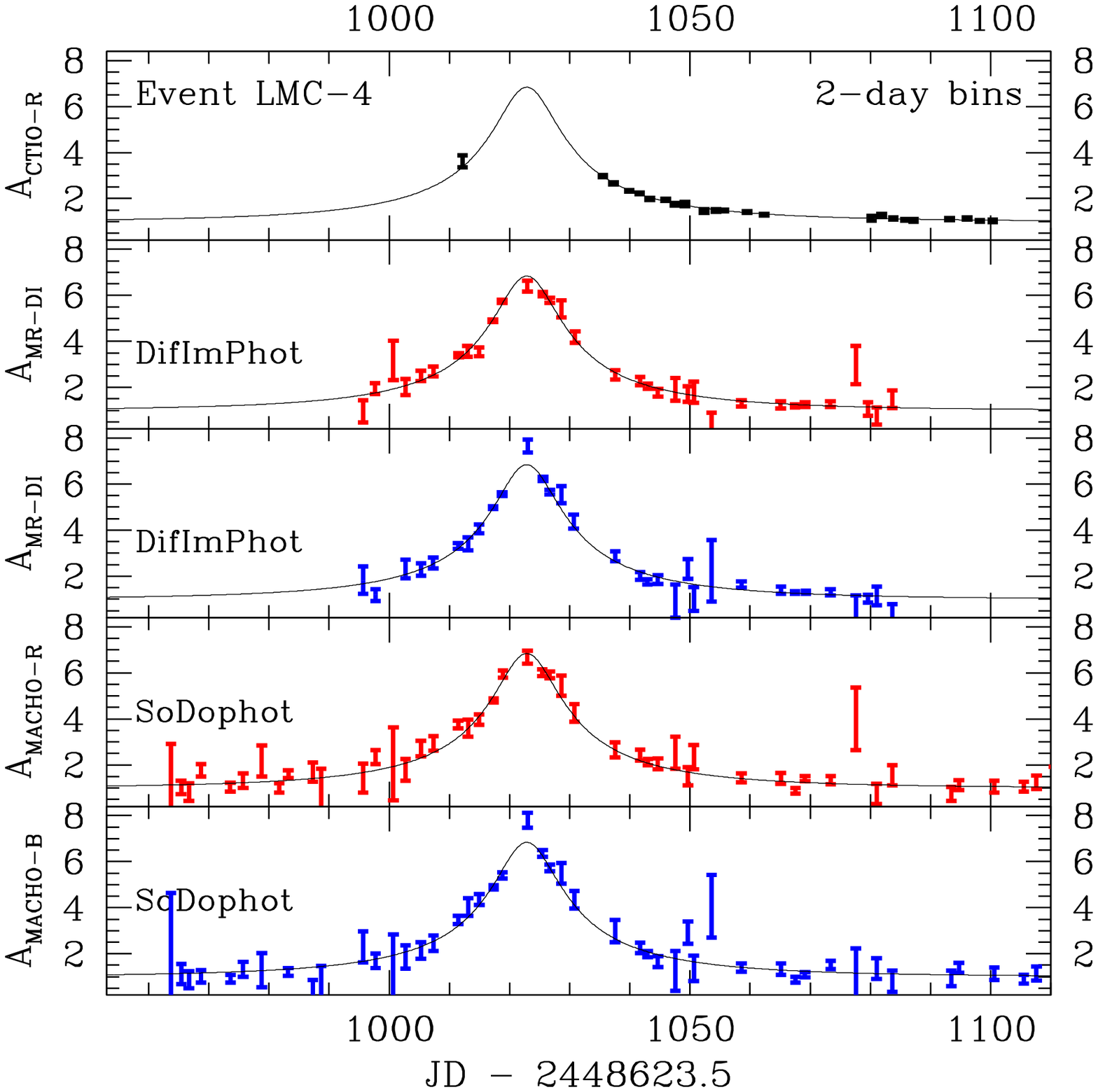}
\caption{
The MACHO and GMAN-CTIO (top panel) follow-up data are shown in
a close-up of the light curve for
microlensing event LMC-4. All data is presented in 2-day bins, and 
the 2-band MACHO data is presented with the original SoDophot photometry 
(bottom 2 panels) as well as photometry from the DifImPhot package 
(2nd and 3rd panels.
\label{fig-lmc4}}
\end{figure}

The full MACHO LMC-4 light curve is shown in Fig.~\ref{fig-lmc4long} and 
Fig.~\ref{fig-lmc4} shows a light curve close-up. The black curve in
each plot is the best fit standard \pac\ light curve, while the fit parameters
and $\chi^2$ values are given in Tables \ref{tbl-fitpar} and \ref{tbl-chi2}. 
The fit $\chi^2$ decreases by only $0.16$ with one additional degree of 
freedom, if the HST constraint on the baseline brightness in the MACHO-blue 
band is released. The data in Fig.~\ref{fig-lmc4long} have been combined 
into 8-day bins and the data shown in Fig.~\ref{fig-lmc4} are displayed in 
2-day bins in order to show how well the data fits the model light curve. 
Since there are as many as 5 observations per day for this event, 
a display without 
binning is easily dominated by the measurements with large error bars, 
and the strength of the signal is easily obscured (as shown in 
\citep{BEL-lmc}). The DifImPhot photometry provides error bars that are 
$\sim 75$\% of the SoDophot error bars, and the fit $\chi^2/{\rm d.o.f.}$ 
values are similar. The CTIO photometry has error bars that average
$\sim 42$\% of the SoDophot error bars with $\chi^2/{\rm d.o.f.} \simeq 1$,
so the DifImPhot and CTIO photometry clearly support the microlensing 
model.

Table \ref{tbl-delchi2} shows $\Delta\chi^2$, which 
is the improvement in the fit $\chi^2$ over a constant brightness model. 
As we should expect from the size of the error bars, the $\Delta\chi^2$ 
values are about a factor of two larger for the DifImPhot data than with 
the original SoDophot data. The $\Delta\chi^2$ contribution from the
CTIO data is smaller, despite the smaller error bars, because of limited
coverage of the CTIO observations.

The one apparent discrepancy between the model and the data is the
MACHO-blue point at the peak of the light curve, a bin which is the 
average of the 6 observations from the night of  1994, Oct.~20, 
and 2 observations from Oct.~21, when most of the night was 
clouded out. This high point is caused by 3 of these 8 unbinned 
measurements, which are between 2-$\sigma$ and 3-$\sigma$ above 
the best fit curve with the SoDophot photometry. 4 of the remaining
measurements from these two days are within $1-\sigma$ of the 
best fit curve, and the remaining point is about 1.8-$\sigma$ below 
the best fit curve. The 3 high points are immediately preceded and
followed by other observations on the same day that are within
1-$\sigma$ of or below the best fit light curve.
For the DifImPhot photometry, the 3 high points
are between 1.5-$\sigma$ and 2.2-$\sigma$ above the best fit curve,
but the point below the curve has moved much closer to the fit curve.
So, although the DifImPhot photometry reduces the scatter, the 
average of the 8 data points is $\sim 3$-$\sigma$ above the curve for
both the SoDophot and DIfImPhot photometry. 

All evidence suggests that these high photometry points are due to
a systematic photometry errors instead of some intrinsic change in
blue-band brightness variation. The high points are not correlated
in time, and they are not reproduced in the MACHO-red band 
despite the fact that the two MACHO passbands have some
overlap in wavelength. This strongly suggests that this high
point is due to a systematic photometry error, and observers
report suggests a possible cause of such an error. The report from  
Oct.~20 began with the following description of the observing conditions:
`Observing conditions fairly poor for most of the night, drifting cloud 
patches and a bright moon. Seeing poor, 2.7"--3.5".' These bright moon 
conditions did not affect any other part of the magnified part of the 
light curve, so we conclude that a systematic photometry error related to the
bright sky background is the most likely cause of the MACHO-blue deviation
at the peak of the light curve.

The source star lies on the main sequence in the HST color-magnitude 
diagram \citep{macho-hstlmc1,macho-hstlmc2}, and its fit R-magnitude 
is within 1--$\sigma$ of the HST 
R-magnitude, so the microlensing model correctly predicts the source star
color. (The uncertainty in the fit $R_M$ magnitude is the quadrature sum 
of the fit error (0.04 mag in this case), the 0.04 mag calibration uncertainty 
\citep{macho-calib}, and the uncertainty in the HST V-magnitude that 
the fit is normalized to (0.03 mag). Table~\ref{tbl-delchi2} shows 
$\Delta\chi^2$, the improvement in $\chi^2$ between the microlens model 
fit and a constant brightness star fit. While the CTIO data 
provide substantially more precise photometry than the Mt. Stromlo data, 
the much better light curve coverage means that most of strength of the 
microlensing signal comes from the Mt. Stromlo data. The addition of the
CTIO data and the DifImphot photometry strengthens the microlensing
signal considerably, from $\Delta\chi^2 = 11,518$ (from the SoDophot
photometry alone) to  $\Delta\chi^2 = 27,973$ (from the DifImphot
photometry and CTIO data), with no indication of a deviation from
the \pac\ light curve (aside from the apparent systmatic error at the
light curve peak). The $\Delta\chi^2$ value for LMC-4 with the CTIO
and DifImPhot photometry ranks it second among on the MACHO LMC
microlensing event candidates, behind LMC-14 \citep{macho-96lmc2} and
in an approximate tie with LMC-1.

\subsection{Event MACHO LMC-13}
\label{event-lmc13}

\begin{figure}
% \plotone{lclmc13_3plongav8.eps}
\plotone{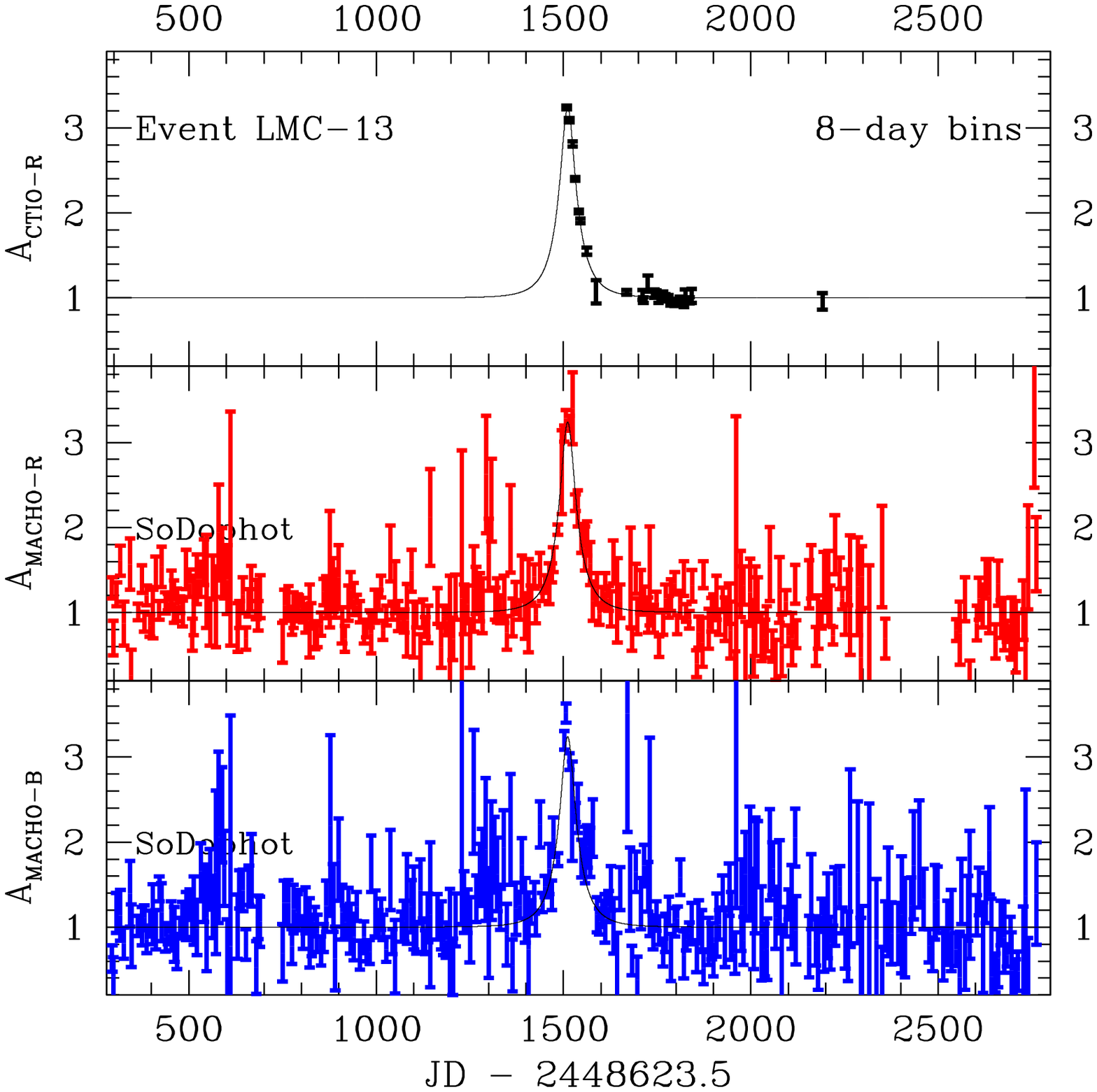}
\caption{
The MACHO and GMAN-CTIO (top panel) follow-up data are shown 
in the full light curve for microlensing event LMC-13
(also known as  MACHO-96-LMC-1). 
The solid line is the best fit light curve constrained to match 
the HST V-band magnitude, and all data is presented in 8-day bins.
\label{fig-lmc13long}}
\end{figure}

\begin{figure}
% \plotone{lc96l1_5pfigav4med.eps}
\plotone{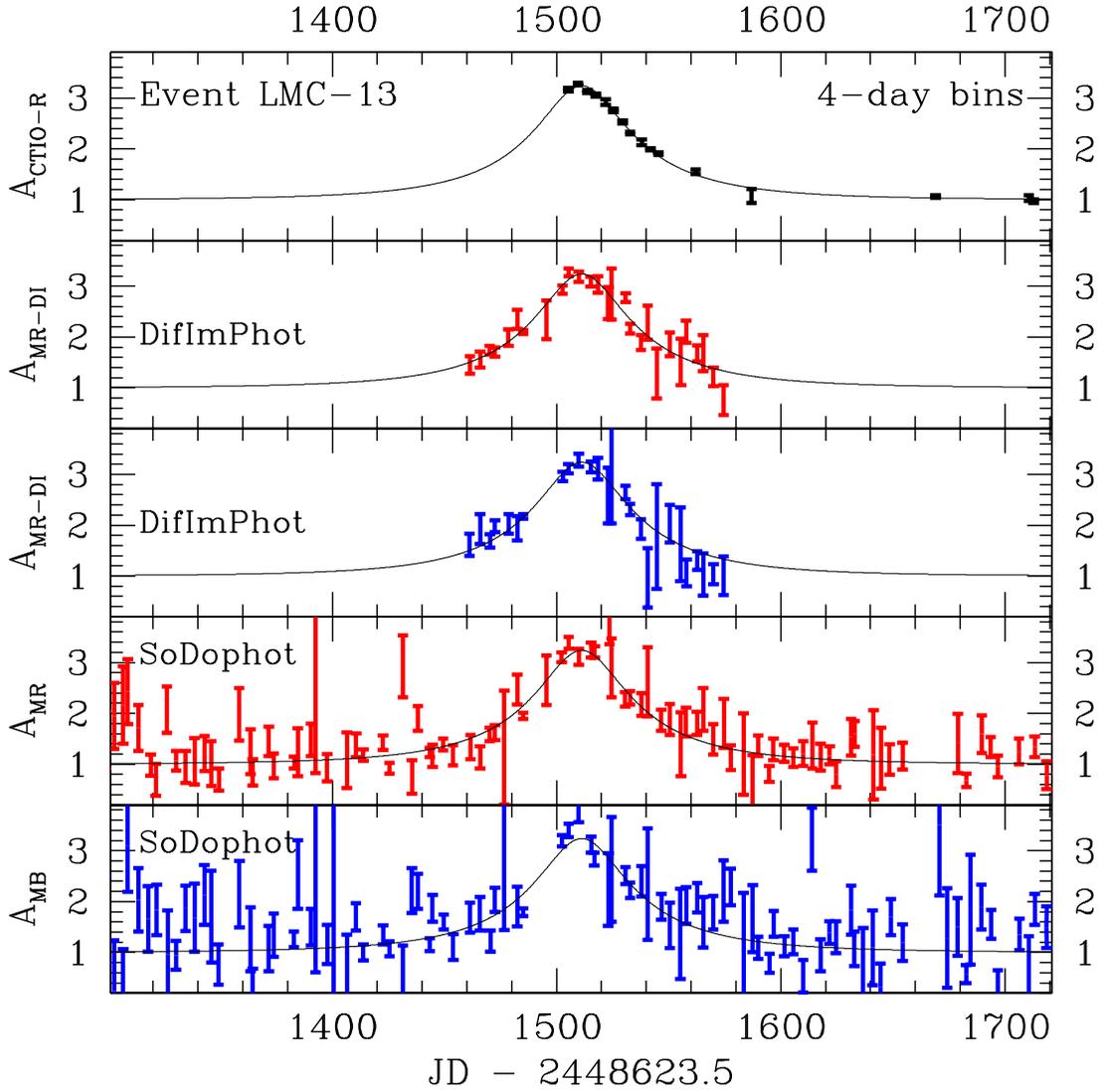}
\caption{
The MACHO and GMAN-CTIO (top panel) follow-up data are shown in
a close-up of the light curve for
microlensing event LMC-13. All data 
is presented in 4-day bins, and the 2-band MACHO data is presented 
with the original SoDophot photometry (bottom 2 panels) as well as 
photometry from the DifImPhot package. For this event, the photometric 
noise has been substantially reduced by the DifImPhot photometry.
\label{fig-lmc13}}
\end{figure}

Figs.~\ref{fig-lmc13long} and \ref{fig-lmc13} show the full light curve
and a close-up of the LMC-13 light curve, which is
displayed in 8 and 4-day bins, respectively. This event shows the
greatest improvement with the DifImphot photometry, with DifImPhot error bars
that are, on average, $\sim 67$\% of the SoDophot error bars. Also, in this
case, CTIO observations were able to begin shortly after the alert, just
prior to peak magnificaiton, and the CTIO data show that the light curve
follows the microlensing model very accurately from peak magnification down to 
to the baseline. Fig.~\ref{fig-lmc13} also reveals some apparent 
photometric errors in the SoDophot MACHO-blue light curve. Apparent 
deviations above the fit curve at the peak and near day 1575 disappear
in the DifImPhot data set as does a low point with small error bars near
day 1485. These systematic errors also appear to interfere with the
blending estimate from the light curve fit. An unconstrained fit predicts
a source star V magnitude that misses the HST value by about 3.4 times
the 1--$\sigma$ uncertainty,
and the fit with the source star magnitude constrained to the HST magnitude
increases the $\chi^2$ value of the fit by 7.08 which corresponds to a
2.7--$\sigma$ discrepancy. 
However, most of this $\chi^2$ is from the MACHO-blue SoDophot
photometry while the MACHO-blue DifImphot photometry has an improved
$\chi^2$ with the HST normalized fit. Thus, we believe that the HST
normalized fit is probably more accurate.

From Table~\ref{tbl-fitpar}, we can see that our fit matches the HST
R-magnitude exactly, and the source star is on the main sequence of the
HST color-magnitude diagram \citep{macho-hstlmc2}. 
While the DifImPhot photometry reduces the photometry
error bars by a factor of 2/3, the CTIO follow-up photometry has error bars that
are, on average a factor of 5 smaller than the original SoDophot photometry of
the MACHO data. The fact that the \pac\ light 
curve shape is confirmed with photometry 5 times more accurate than the
MACHO SoDophot photometry that the event identification was based on
provides strong support for the microlensing interpretation of this event.

\subsection{Event MACHO LMC-15}
\label{event-lmc15}

\begin{figure}
% \plotone{lclmc15_3plongav8.eps}
\plotone{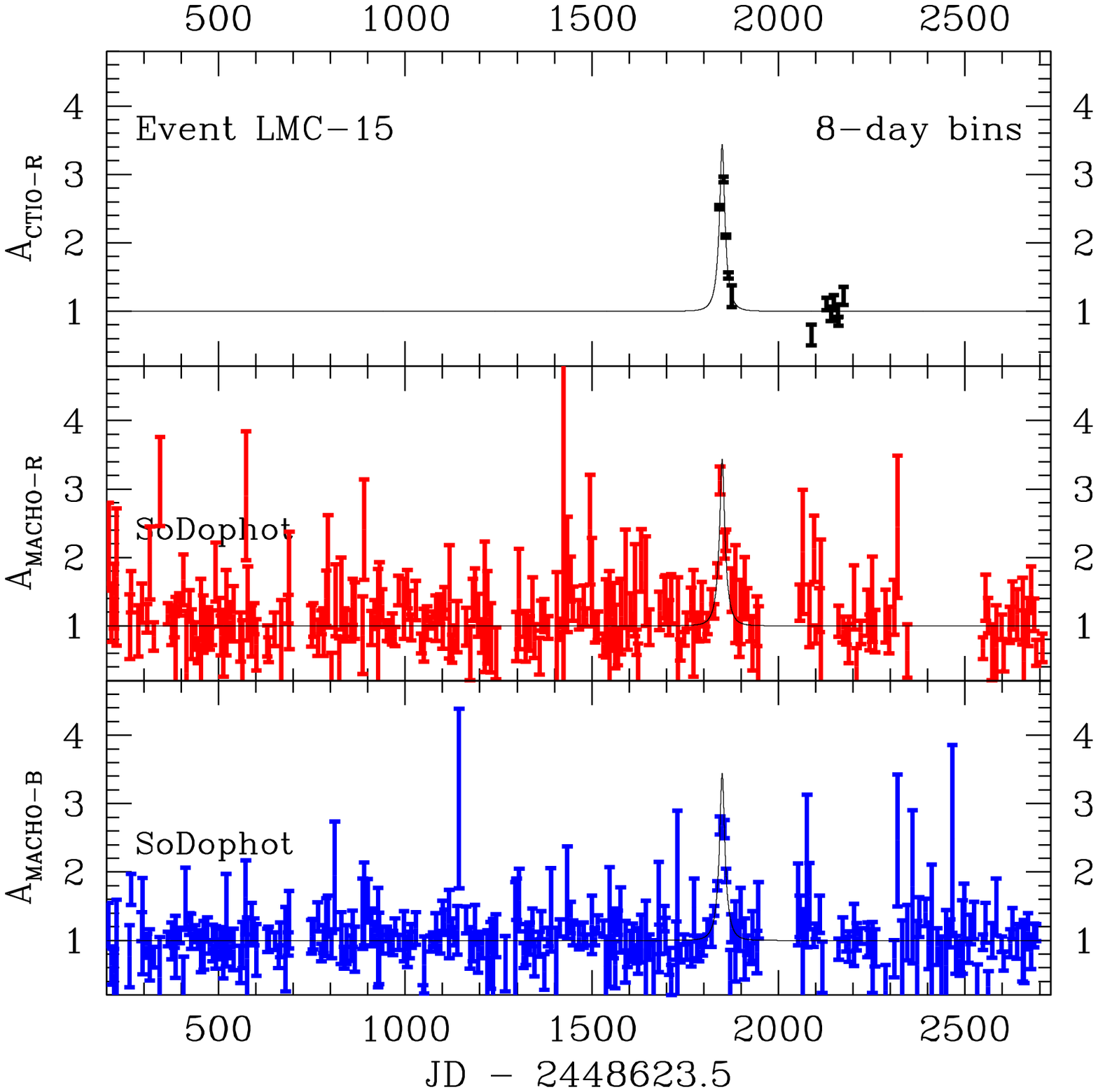}
\caption{
The MACHO and GMAN-CTIO (top panel) follow-up data are shown 
in the full light curve for microlensing event LMC-15
(also known as  MACHO-97-LMC-1). 
The solid line is the best fit light curve constrained to match 
the HST V-band magnitude, and all data is presented in 8-day bins.
\label{fig-lmc15long}}
\end{figure}

\begin{figure}
% \plotone{lc97l1_5pfigav2.eps}
\plotone{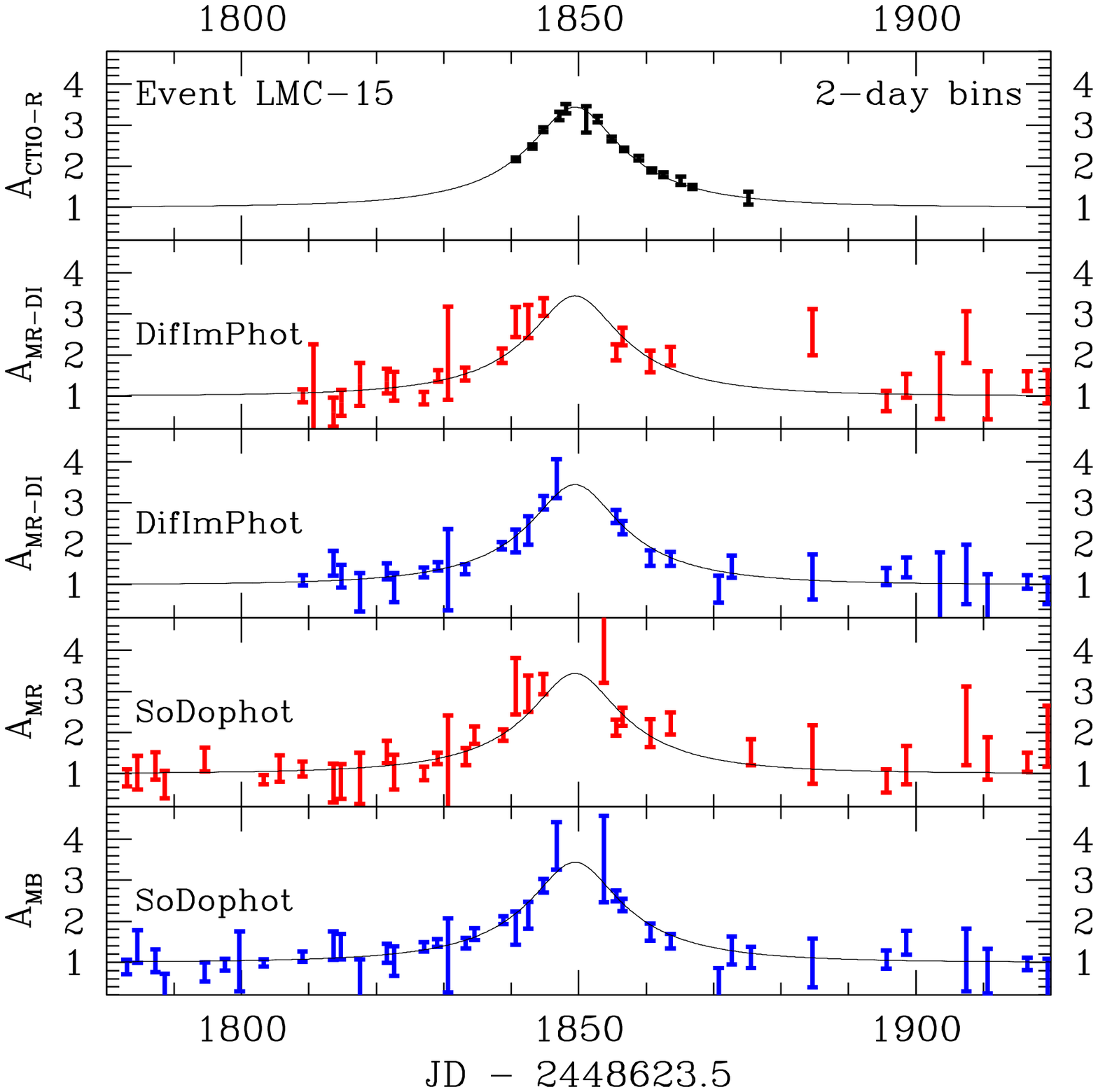}
\caption{
The MACHO and GMAN-CTIO (top panel) follow-up data are shown in
a close-up of the light curve for
microlensing event LMC-15. All data 
is presented in 2-day bins, and the 2-band MACHO data is presented 
with the original SoDophot photometry (bottom 2 panels) as well as 
photometry from the DifImPhot package.
\label{fig-lmc15}}
\end{figure}

The full light curve of MACHO LMC-15 is shown in Fig.~\ref{fig-lmc15long} in
8-day bins, a close-up is shown in Fig.~\ref{fig-lmc15},
displayed in 2-day bins. This event was the lowest signal--to--noise
event to pass the stricter ``criterion A" of the MACHO LMC 5.7 year analysis.
Fortunately, the event was identified well before peak magnification, and
very good light curve coverage was obtained from CTIO that filled the
gaps in the data from Mt.~Stromlo. Furthermore, the error bars on the
CTIO measurements are, on average, a factor of 4 smaller than the MACHO-red
error bars and a factor of 4.8 smaller than the MACHO-blue error bars,
while the CTIO $\chi^2/{\rm d.o.f.}$ is quite reasonable. So, the CTIO
data provides strong confirmation of the microlensing hypothesis.
The DifImPhot
photometry, on the other hand, provides only a modest improvement over
SoDophot with error bars that are $\sim 85$\% of the SoDophot
error bars in the MACHO-red band and $\sim 95$\% of the SoDophot
error bars in the MACHO-blue band.

The fit parameters presented Table~\ref{tbl-fitpar} indicate that the fit
R-band magnitude matches the HST value to better than $0.5\sigma$,
and the source star is also on the HST main sequence
\citep{macho-hstlmc2}. Also,
the constraint that the MACHO-blue magnitude be fixed to match the
HST V-band magnitude only causes $\chi^2$ to increase by $0.39$, so
the fit V-band magnitude also matches the HST V-band magnitude to better
than 1--$\sigma$.

\subsection{Counterexample: Event MACHO LMC-23}
\label{event-lmc23}

\begin{figure}
% \plotone{lclmc23_4pfigav2.eps}
\plotone{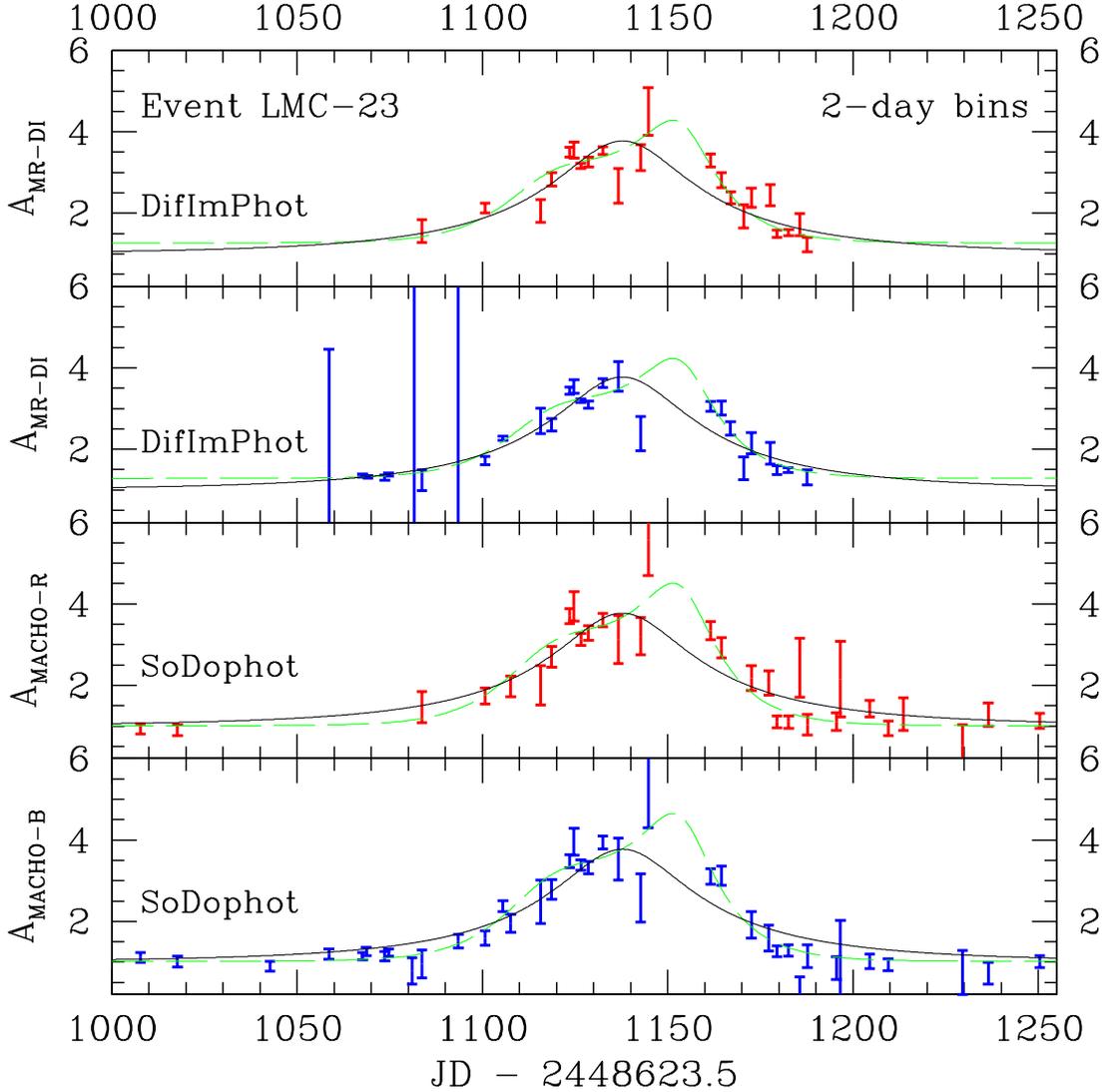}
\caption{
The MACHO data are shown in a close-up of the light curve of
pseudo-microlensing event LMC-23, which 
has recently been observed to have a 2nd brightening episode which 
makes the microlensing interpretation extremely unlikely. The data is 
presented in 2-day bins, with both the original SoDophot photometry 
(bottom 2 panels) as well as photometry from the DifImPhot package.
The dashed green curve is a microlensing parallax fit, which provides 
a substantially better fit to the data (but is also unlikely to be correct).
\label{fig-lmc23}}
\end{figure}

Microlensing event candidate MACHO LMC-23 was not detected by the 
MACHO alert system, so there is no CTIO data for it. Some DifImphot
photometry does exist, with error bars that average $\sim 60$\% of the
SoDophot error bars. It is obvious from Fig.~\ref{fig-lmc23}
that there are deviations from the standard \pac\ light curve fit. Both the
MACHO-red and blue data are brighter than the standard \pac\ light
curve fit near days 1123 and 1163, and this occurs for both the
SoDophot and DifImPhot photometry. All 4 data sets appear to be below
the fit curve near day 1180, as well. A microlensing parallax model
(the dashed green line) provides a much better fit, and seems to fit the
deviations at days 1163 and 1180. There still appears to be a deviation 
near day 1123, however.

These qualitative observations are also reflected in the fit $\chi^2$ values
shown in Table~\ref{tbl-chi2}. For the standard \pac\ light curve fit, the
peak $\chi^2/{\rm d.o.f.}$ values are quite large ranging from 3.229 for the
MACHO-blue SoDophot data to 5.814 for the blue DifImPhot data. For the
microlensing parallax model fit (labeled {\it LMC-23-p}), the 
$\chi^2/{\rm d.o.f.}$ values drop to reasonable values for the SoDophot
photometry (except perhaps the peak $\chi^2/{\rm d.o.f.}$ value for the
MACHO-red data), but the DifImPhot values remain high at 2.434 and
3.012, respectively, for MACHO-red and blue. This suggests that neither
model is likely to be correct, but it is possible that a binary lens model might
do better.

Another problem is that the fit R-band magnitudes do not match the HST
R-band magnitude. They are off by $V-R \approx 0.3$, which is
more than 4--$\sigma$ for both the standard and parallax fit.  The
source star is on the sub-giant branch of the HST color magnitude diagram
\citep{macho-hstlmc2}, but the color at maximum brightness is more
consistent with a main sequence star.
On the other hand, the apparent redness of the source star 
could possibly be explained by blending in
the HST images if the lens star were visible and redder than the source,
as was the case for event 5 \citep{macho-hstlmc5}.

However, an additional complication is the fact that the EROS Collaboration
has observed a subsequent brightening of this star approximately 2500 days
after the one observed by MACHO \citep{glicens-hawaii}. 
Multiple brightening episodes can 
occur in binary lens or binary source microlensing events, such as 
MACHO 96-BLG-4 \citep{macho-binaries}, if the binary lens or source
separation is large. However, the probability of detecting such an event
decreases in proportion to the binary separation because alignment
of both members of the binary with the single lens or source decreases
with their separation. Thus, while it might be possible to find a model
that could explain this event as lensing, it would require several
highly improbable occurrences for this single event: two different
types of exotic lensing to explain the light curve shape in the MACHO
data as well as the brightening seen by EROS, plus a bright lens
(or unlikely HST blend by chance superposition). It is much more
likely that MACHO LMC-23 is a variable star. This is currently the
only MACHO LMC microlens candidate that we have any reason to
suspect is not an actual microlensing event.

\section{Confrontation of BEL Event Classification with Additional Data}
\label{confront}

The classification of the MACHO LMC events by the MACHO team and 
BEL are listed in Table~\ref{tab-class}. This table also provides
a ``confirmation" column which indicates the implication of additional
data for the event classification. Items that support the microlensing
interpretation are listed in bold face, and items that tend to 
contradict the microlensing interpretation are presented in
italics. The MACHO verdict refers to the classification given 
in A00. Events listed as $\mu$lens-A and $\mu$lens-B are
candidate microlensing events that passed the strict criteria A,
or only the looser B criterion. Event 22 did pass criteria B, but
a spectrum taken with the MSSSO 2.3m telescope revealed that
the source was an emission line galaxy at $z = 0.233$ 
(T.~Axelrod, private communication).

Table~\ref{tab-class} indicates that
our expectation that the BEL event selection method is unreliable has 
been borne out by our new data and analysis. Of the 4 events that 
we have analyzed, the only one that BEL classify as microlensing is
event 23, which is most likely a variable star. The other 3 events they
classify as non--microlensing, but the new data confirms the
microlensing interpretation, again contradicting BEL.
These 4 events probably provide the best test of the BEL method, 
because these data had not been previously published, except in
a PhD thesis \citep{becker-thesis}, and so, the additional data
were probably not seen by BEL when they were designing their
method. In contrast, the original MACHO analysis \citep{macho-lmc5.7}
is confirmed for the three events with additional photometric data
during the microlensing event, but not for event 23.

The complete MACHO analysis identifies 25 events as either 
microlensing candidates or supernovae. (The events are numbered
1, 4--27, as events 2 and 3 from the first year analysis 
\citep{macho-lmc1} do not make the 5--year cuts.) Of these
25 events, MACHO classifies 16 as microlensing candidates,
8 as background supernovae, and 1 as either an AGN or a
peculiar background supernova. The BEL analysis identifies
10 of these events as microlensing, 5 as supernovae, and 9
as neither. Event 9, the LMC binary lens event 
\citep{macho-lmc9,macho-binaries}, is not considered by BEL.
The BEL analysis agrees with MACHO in the classification of 7
events as microlensing and 4 events as supernovae. We will
now examine these classifications in light of other, non--light 
curve data.

Of the 7 events identified by both MACHO and BEL as microlensing,
event 1 has a follow-up spectrum consistent with the microlensing 
interpretation \citep{massimo-lmc1}, and event 5 has been confirmed 
by HST observations that identified the lens star in the Milky Way disk 
and provided the first complete solution for a microlensing event
\citep{macho-hstlmc5,dck-lmc5,lmc5-mass,spitz-lmc5}.
Event 14 was confirmed with high signal-to-noise
GMAN follow-up data which also enabled measurement of the microlensing
``xallarap" effect (light curve oscillations due to the orbit of a 
binary source star) and the determination that the lens is close to 
the LMC \citep{macho-96lmc2}. We argue in Sec.~\ref{conclude} that
events 1 and 25 are very likely to be microlensing, as well, but
we have seen that event 23 is probably a 
variable star although it was identified as a microlensing candidate by
both MACHO \citep{macho-lmc5.7} and BEL. Thus, 4 of these 7
events can be considered to be confirmed, and 1 is rejected.
 
Three events are identified by BEL, but not MACHO,  as microlensing:
events 10, 22, and 24. They all have highly asymmetric light curves that suggest 
a non-microlensing origin, although in the case of event 22, the 
asymmetry can be fit by a microlensing parallax model with plausible 
parameters. The MACHO photometry of event 10 is a near perfect fit 
to a supernova type Ia light curve, and so this event is clearly a 
background supernova. It is near the blue end of the SN type 1a
color distribution, and this is probably why BEL's SN test fails.
Event 24 is also fit much better by a supernova 
type Ia light curve than by standard microlensing, so it was also 
rejected as a microlensing candidate. The quality of the type Ia fit 
for event 24 is poor, however, so it is more likely that this event 
is a type II supernova.  The source locations for each of these events 
correspond to background galaxies. For event 24, the galaxy is visible 
in the MACHO images, while the source ``star" for event 10 is revealed 
to be a compact galaxy in HST images \citep{macho-hstlmc1}.
The source for LMC-22 
appears slightly extended in a higher resolution CTIO 4m frame, and 
a spectrum of the source indicates that it is probably an active galaxy.
Thus, the microlensing 
interpretation is rejected for all three of these events classified by 
BEL, but not MACHO, as microlensing. 

Of the 9 events classified by BEL as neither microlensing nor
supernovae, MACHO classifies 2 as supernovae, and 7 as
microlensing (events 4, 7, 8, 13, 15, and 18). In 
Secs.~\ref{event-lmc4}--\ref{event-lmc15}, we have shown that
the microlensing interpretations of 3 of these events (4, 13, and 15)
are strongly confirmed by the additional photometry we have
presented, but there exists 
no follow-up data that sheds light on the other events. 
However, event 8 is the lone event in the center of the LMC bar, where
the LMC self-lensing probability is highest \citep{man-lmc-mod},
so it seems likely that this event is microlensing, as well.
Thus, the BEL classification fails for 3 of these events and
is likely to fail for one other.

It seems clear that the only category in which BEL's
event classification scheme appears to succeed, more often
than not, is when it agrees with the MACHO classification.
We have shown that their identification of events as
``non-microlensing"  fails for the 3 events with new data presented
in this paper, and in most cases where they disagree with MACHO,
additional data strongly suggests that BEL is wrong. We can only
conclude that BEL's selection method does not
provide any help in separating real microlensing events
from other types of variability that mimic microlensing.
However, this does not mean that neural networks cannot 
improve upon the MACHO analysis. Instead, as we have argued
in Sec.~\ref{nnet}, that BEL's method fails because of the
weak set of statistics that they have relied upon. A neural network
that uses the more powerful light curve fitting based statistics
used by MACHO might indeed offer an improvement. 

\section{Discussion and Conclusions}
\label{conclude}

The MACHO Project's 5.7 year LMC analysis has identified a set of
16 candidate LMC microlensing events after removing the background
supernovae. In this paper, we have analyzed additional photometry
from GMAN follow-up observations with the CTIO 0.9m telescope as
well as DifImPhot photometry of the original Mt.~Stromlo MACHO
observations for three of these events: MACHO LMC-4, 13 and 15. 
This additional photometry has errors that are smaller
than the original MACHO photometry by a factor of 1.1-1.5 (for 
DifImPhot) and of 2.4-5 (for the GMAN CTIO photometry), and 
improves the light curve time coverage for these events. All three light
curves with the improved photometry are well fit by a standard \pac\
microlensing model.

For two of the three light curves, the fit accurately predicts both the
V and R--band magnitudes from high resolution HST observations
of the microlensed source star. For the
other event, LMC-13, the unconstrained \pac\ fit predicts a V--band 
magnitude that differs by the HST magnitude by 2.7--$\sigma$, but
the difference between the DifImPhot and SoDophot photometry of the
Mt.~Stromlo data suggests that this may be affected by low level systematic
photometry errors. A \pac\ light curve fit constrained to the HST V
magnitude accurately predicts the HST R magnitude for event LMC-13.
Thus, it is fair to say that the additional photometry provides a strong
confirmation of the microlensing interpretation of these three events.

In addition to these events, there are several other of the MACHO
LMC microlensing candidates for which the microlensing interpretation
seems very secure. Event 14 has also been confirmed with GMAN 
photometry that provides evidence of the ``xallarap" effect 
\citep{macho-96lmc2}, and Event 5 is the first microlensing event
in which the lens has been detected 
\citep{macho-hstlmc5, dck-lmc5, spitz-lmc5}
and has been completely solved yielding the mass, transverse
velocity and distance of the lens \citep{lmc5-mass}. The consistency
of the microlensing mass determination with the lens star brightness
and colors provides a strong confirmation of the microlensing model
for this event.

Events LMC-1 and LMC-25 are the two events which clearly have red 
clump giant source stars 
\citep{massimo-lmc1,macho-lmc5.7,macho-hstlmc2}, 
and such stars have
not exhibited the type of variability that could be confused with 
microlensing \citep{blue-var} as observed in EROS LMC-1 and 2
and MACHO LMC-2 and 23 
\citep{macho-lmc2,eros2-var,eros12-spec,eros-lmc-tau,glicens-hawaii}.
Finally, event MACHO LMC-9 is well fit by a binary lens model
\citep{macho-lmc9,macho-binaries} and has a light curve shape
that is different from other known forms of variability with the 
very rapid brightness variation that is characteristic of a 
binary lens caustic crossing.

So, out of the 16 MACHO LMC microlensing candidates, 5 have
been confirmed with additional data, and another 3 appear
very likely to be microlensing. Of the 5 events that have been
confirmed with additional data, 4 have been selected by a completely
independent method--their discovery by the MACHO alert system. All
4 of these events have been confirmed to be microlensing, which
suggests that most of the MACHO LMC microlensing candidates
are indeed due to microlensing. Nevertheless, the example of
MACHO LMC-23 suggests that some caution is warranted regarding the
interpretation of 7 remaining MACHO LMC microlensing events
that have not yet been confirmed.

It is instructive to consider the results of \citet{man-lmc-mod} who
have done detailed predictions of the properties of LMC
self-lensing events based upon some of the latest models of the
LMC. (Their classifications are listed in Table~\ref{tab-class}.)
They identify 4 of the MACHO events that are good LMC
self-lensing candidates: events 6, 8, 13, and 14. A straight--forward
extension of their analysis also adds the binary lens event, 9, to
this list. This identification is strengthened by independent 
light curve analyses that suggest that events 9 and 14 both
have lenses residing in the LMC \citep{macho-lmc9,macho-96lmc2}. 
Of the remaining 10 events, the event 5 lens star
has been shown to be a disk star. It is most likely a member of 
the thick disk, but thin disk membership cannot be ruled out
\citep{lmc5-mass}. Event 20 is also a disk lens candidate
because it appears to be blended with a star that is much
redder than other LMC stars of similar magnitude 
\citep{macho-lmc5.7}. (If event 20 is also a disk lens this would
give us a total of 2 disk events when a total of 0.75 are expected
from the combined thin and thick disks. The Poisson probability 
to detect 2 or more disk events when 0.75 are expected is 17\%.)
This leaves 8 events as halo lens candidates: events 1, 4, 7, 15, 
18, 21, 25 and 27. Two of these halo lens candidates (4 and 15) are 
confirmed by the photometry and light curve fitting presented
here, and two others (1 and 25) are considered ``solid" events
because of their red clump giant source stars. Thus, there is
little difference between the quality of the different categories of
events. Thus, the ``confirmed" events appear to be distributed
about evenly among the disk, halo, and LMC--self
lensing candidate events. 

Our qualitative conclusion is that
the puzzle of the excess LMC lensing events remains unresolved.
Our knowledge of the Galactic disk and LMC suggest that most
of the lenses must reside in a previously undiscovered population,
and the Milky Way halo is the natural location for this, since the
halo's mass is largely unaccounted for. But such a population
presents a number of problems if it is composed of known objects
like white dwarfs. The leading alternative explanation is that
LMC self-lensing dominates, but this possibility gets no significant
observational support from detailed LMC models or from the
properties of the microlensing events themselves. (See
\citet{sahu03} for a different point of view.)

A resolution of this puzzle will probably require additional data
so that the distance to a representative sample of LMC lensing
events can be determined. This could come from the microlensing
key project for the SIM mission \citep{sim}, or from a Deep 
Impact mission extension (DIME) \citep{dime}, which would
use the 30cm telescope on the Deep Impact spacecraft to
make microlensing parallax observations from a Heliocentric
orbit \citep{gould-parsat}, if it is approved. 
Microlensing experiments towards
other lines of site could also shed some light on this
issue \citep{m31-mega,m31-tau,m31-pa}. In addition, the SuperMACHO 
\citep{becker-sm} and MOA-II \citep{moa2-hawaii} surveys expect to
substantially increase the detection rate of LMC microlensing
events in order to provide alerts for SIM and DIME and to
measure the spatial variation of the microlensing
optical depth across the face of the LMC.

\acknowledgments

D. P. B. was supported by
grants AST 02-06189 from the NSF and NAG5-13042 from
NASA.

\clearpage


\clearpage

\begin{deluxetable}{lllllllcccr}
\tabletypesize{\scriptsize}
\tablecaption{Microlensing Fit Parameters \label{tbl-fitpar}}
\tablewidth{0pt}
\tablehead{
\colhead{Event} & \colhead{$f_{\rm MR}$} & \colhead{$f_{\rm MB}$} &
\colhead{$f_{\rm CTIO}$} & \colhead{$V_{\rm M}\equiv V_{\rm HST}$} &  
\colhead{$R_{\rm M}$} &  \colhead{$R_{\rm HST}$} & 
\colhead{$t_0$ (MJD)} & \colhead{$u_0$} & 
\colhead{$t_E$ (days)} & \colhead{${\chi^2\over {\rm(dof)}}$} 
}
\startdata
LMC-4 & $0.384(7)$ & $0.335$ & $0.598(16)$ & $21.33(3)$ & $21.15(6)$ & $21.09(3)$ &
$1022.86(6)$ & $0.147(2)$ & $39.5(7)$ & $1.418$ \\
LMC-13 & $0.66(2)$ & $0.564$ & $0.77(2)$ & $21.76(3)$ & $21.38(6)$ & $21.38(3)$ &
$1511.2(2)$ & $0.320(7)$ & $66.0(1.3)$ & $1.177$ \\
LMC-15 & $0.73(6)$ & $0.755$ & $0.65(3)$ & $21.18(3)$ & $21.10(9)$ & $21.07(3)$ &
$1849.4(1)$ & $0.300(12)$ & $22(1)$ & $0.979$ \\
{\it LMC-23} & $0.42(2)$ & $0.508$ & -- & $21.05(3)$ & ${\it 20.92(6)}$ & $20.64(3)$ &
$1137.8(5)$ & $0.272(8)$ & $70(2)$ & $1.730$ \\
{\it LMC-23-p} & $0.40(2)$ & $0.508$ & -- & $21.05(3)$ & ${\it 20.95(6)}$ & $20.64(3)$ &
$1135.8(4)$ & $-2.48(16)$ & $16.2(8)$ & $1.410$ \\
 \enddata
\tablecomments{ The columns of this table list the fit parameters for each event. $f_{\rm MR}$,
$f_{\rm MB}$, and $f_{\rm CTIO}$ are the blend fractions for the MACHO red and blue, and
CTIO data sets. $V_{\rm M}$, $R_{\rm M}$, $V_{\rm HST}$, and $R_{\rm HST}$ are the
calibrated V and R band source star brightnesses for the MACHO \citep{macho-calib} 
and HST photometry.$t_0$, $u_0$ and $t_E$ are the standard \pac\ microlensing
light curve fit parameters. $t_0$ and $u_0$ are the time of peak magnification and the
impact parameter, given in units of the Einstein ring radius. $t_E$ is the Einstein
radius crossing time.
}
\end{deluxetable}

\begin{deluxetable}{llrrrrrrrrrr}
% \tabletypesize{\scriptsize}
\tabletypesize{\small}
\tablecaption{Microlensing Fit $\chi^2$ Values \label{tbl-chi2}}
\tablewidth{0pt}
\tablehead{ & & 
\multispan{2}\hfil {MACHO-R}\hfil & \multispan{2}\hfil {MACHO-R-DI}\hfil & 
\multispan{2}\hfil {MACHO-B}\hfil & \multispan{2}\hfil {MACHO-B-DI}\hfil &
\multispan{2}\hfil {CTIO-R}\hfil \\
\colhead{Event} & \colhead{} & 
\colhead{$N_{\rm data}$} & \colhead{${\chi^2\over {\rm(dof)}}$} &
\colhead{$N_{\rm data}$} & \colhead{${\chi^2\over {\rm(dof)}}$} &
\colhead{$N_{\rm data}$} & \colhead{${\chi^2\over {\rm(dof)}}$} &
\colhead{$N_{\rm data}$} & \colhead{${\chi^2\over {\rm(dof)}}$} &
\colhead{$N_{\rm data}$} & \colhead{${\chi^2\over {\rm(dof)}}$} 
}
\startdata
LMC-4 & full & 
$730$ & $1.368$ & $85$ & $1.355$ & $728$ & $1.522$ & $78$ & $1.742$ & $134$ & $0.989$ \\
& peak & 
$92$ & $1.573$ & $85$ & $1.355$ & $91$ & $1.430$ & $78$ & $1.742$ & $107$ & $1.072$ \\
LMC-13 & full & 
$1074$ & $1.256$ & $84$ & $1.376$ & $1176$ & $1.196$ & $84$ & $1.026$ & $332$ & $0.937$ \\
& peak & 
$165$ & $1.056$ & $84$ & $1.376$ & $171$ & $1.557$ & $84$ & $1.026$ & $182$ & $0.925$ \\
LMC-15 & full & 
$475$ & $0.984$ & $47$ & $1.272$ & $586$ & $0.914$ & $49$ & $1.067$ & $60$ & $1.303$ \\
& peak & 
$29$ & $1.144$ & $25$ & $1.607$ & $32$ & $0.752$ & $28$ & $0.694$ & $39$ & $0.876$ \\
{\it LMC-23} & full & 
$389$ & $1.498$ & $24$ & $4.846$ & $382$ & $1.495$ & $30$ & $5.814$ & $0$ & -- \\
& peak & 
$31$ & $3.413$ & $24$ & $4.846$ & $41$ & $3.229$ & $30$ & $5.814$ & $0$ & -- \\
{\it LMC-23-p} & full & 
$389$ & $1.364$ & $24$ & $2.434$ & $382$ & $1.284$ & $30$ & $3.012$ & $0$ & -- \\
& peak & 
$31$ & $1.688$ & $24$ & $2.434$ & $41$ & $1.310$ & $30$ & $3.012$ & $0$ & -- \\
 \enddata

\end{deluxetable}

\begin{deluxetable}{lrrrrrrr}
% \tabletypesize{\scriptsize}
\tabletypesize{\small}
\tablecaption{$\Delta\chi^2$ Values \label{tbl-delchi2}}
\tablewidth{0pt}
\tablehead{ \colhead{Event} & \colhead{MACHO-R} & \colhead{MACHO-R-DI} &
\colhead{MACHO-B} & \colhead{MACHO-B-DI} & \colhead{CTIO-R} &  \colhead{CTIO-B} & 
\colhead{total}
}
\startdata
LMC-4 & $5540.5$ & $11664.2$ & $5977.4$ & $11687.8$ & $4621.3$ & -- & $27973.3$ \\
LMC-13 & $2037.0$ & $2377.6$ & $1419.2$ & $1543.8$ & $17523.1$ & -- & $21444.5$ \\
LMC-14 & $5229.8$ & $6328.6$ & $10860.0$ & $13868.0$ & $69274.8$ & $64081.7$ & $153553.0$ \\
LMC-15 & $254.3$ & $234.2$ & $554.0$ & $437.8$ & $2204.0$ & -- & $3012.2$ \\
 \enddata
\end{deluxetable}

\begin{deluxetable}{rcccc}  % 4 columns.
\tablecaption{Event Classification \label{tab-class} }
\tablewidth{0pt}
\tablehead{
%% Use a footnote to explain numbering.
%% The qquads widen the table.
\colhead{Event} &
\colhead{MACHO Verdict} &
\colhead{BEL Verdict} &
\colhead{confirmation} &
\colhead{Mancini et al lens type} 
}  % end header.
\startdata

  1 & $\mu$lens-A & $\mu$lens & {\bf clump giant} & non-LMC  \\
  4 & $\mu$lens-A & variable & {\bf CTIO+DIP phot.} & non-LMC \\
  5 & $\mu$lens-A & $\mu$lens & {\bf lens ID} & MW-disk \\
  6 & $\mu$lens-A & $\mu$lens & -- & LMC \\
  7 & $\mu$lens-A & variable & -- & non-LMC \\
  8 & $\mu$lens-A & variable & -- & LMC  \\
  9 & $\mu$lens-B & -- & {\bf caustic-binary} & LMC \\
 10 & SN & $\mu$lens & {\it HST: galaxy} & -- \\
 11 & SN & SN & {\it HST: galaxy} & -- \\
 12 & SN & SN & {\it HST: galaxy} & -- \\
 13 & $\mu$lens-A & variable & {\bf CTIO+DIP phot.} & LMC \\
 14 & $\mu$lens-A & $\mu$lens & {\bf CTIO+DIP phot.} & LMC \\
 15 & $\mu$lens-A & variable & {\bf CTIO+DIP phot.} & non-LMC \\
 16 & SN & -- & {\it CTIO: galaxy} & -- \\
 17 & SN & SN & {\it CTIO: galaxy} & -- \\
 18 & $\mu$lens-A & variable & -- & non-LMC \\
 19 & SN & SN & {\it CTIO: galaxy} & -- \\
 20 & $\mu$lens-B & SN & -- & -- \\
 21 & $\mu$lens-A & $\mu$lens & -- & non-LMC  \\
 22 & B & $\mu$lens & {\it MSSSO: galaxy} & -- \\
 23 & $\mu$lens-A & $\mu$lens & {\it variable} & -- \\
 24 & SN & $\mu$lens & {\it MACHO: galaxy} & -- \\
 25 & $\mu$lens-A & $\mu$lens & {\bf clump giant} & non-LMC \\
 26 & SN & variable & -- & -- \\
 27 & $\mu$lens-B & variable & -- & -- \\

\enddata
\tablecomments{The event classification results of MACHO and BEL are
compared to the results of additional data that can confirm or reject
each event. Confirmed microlensing events have bold face entries in the
confirmation column, and rejected microlensing candidates have entries
in italics}
\end{deluxetable}

%% Text for table notes should follow after the \enddata but before
%% the \end{deluxetable}. Make sure there is at least one \tablenotemark
%% in the table for each \tablenotetext.

\end{document}